\documentclass[prl,aps,showpacs,floatfix,nofootinbib,twocolumn,letterpaper,reprint]{revtex4}
\usepackage{epsfig}
\usepackage{amssymb}
\usepackage{amsmath}
\usepackage{amsfonts}
\usepackage{color,graphicx}
\usepackage{bm}

\newcommand{\ssec}[1]{\emph{#1}.---}

\begin{document}
\title{Nuclear deformation at finite temperature
}
\author{Y. Alhassid,$^{1}$ C.N.~Gilbreth,$^{1}$ and G.F.~Bertsch$^{2}$ }
\affiliation{$^{1}$Center for Theoretical Physics, Sloane Physics
Laboratory, Yale University, New Haven, CT 06520\\
$^{2}$Department of Physics and Institute of Nuclear Theory,
Box 351560\\ University of Washington, Seattle, WA 98915}
\date{\today}
\def\lb{\langle}
\def\rb{\rangle}
\def\ni{\noindent}
\def\be{\begin{equation}}
\def\ee{\end{equation}}
\def\sumk{\sum_k}
\def\ad{a^\dagger_k}
\def\adb{a^\dagger_{\bar k}}
\def\a{a_k}
\def\ab{a_{\bar k}}
\def\Tr{{\rm Tr}}
\def\tr{{\rm tr}}
\def\Re{{\rm Re\,}}

\begin{abstract}
Deformation, a key concept in our understanding of heavy nuclei, is based on a mean-field description that breaks the rotational invariance of the nuclear many-body Hamiltonian. 
We present a method to analyze nuclear deformations at finite temperature in
a framework that preserves rotational invariance. The auxiliary-field Monte-Carlo method is used to generate the statistical ensemble and calculate the probability distribution associated with the
quadrupole operator.  Applying the technique to nuclei in the rare-earth
region, we identify model-independent signatures of deformation and find
that deformation effects persist to higher temperatures than the  spherical-to-deformed shape phase-transition temperature of mean-field theory.   
\end{abstract}


\maketitle

\ssec{Motivation}   Mean-field theory is a useful method for studying correlated many-body systems. However, it often breaks symmetries, making it difficult to compare its results with physical spectra that preserve these symmetries.   In addition, although mean-field theory often predicts sharp phase transitions at finite temperature, they are washed out in finite-size systems. The challenge is to find tools to study the properties of finite-size systems within a framework that preserves the underlying symmetries while also allowing calculation of the quantities that describe symmetry breaking in mean-field theory. 

In nuclear physics, this issue is especially important in the understanding of heavy deformed nuclei, which are of wide experimental and theoretical interest. The current theory of these nuclei is based on self-consistent mean-field (SCMF) theory, which predicts both spherical and deformed ground states~\cite{su3} depending on the nucleus.  SCMF is a  convenient tool to study their
intrinsic structure but it breaks rotational invariance, a prominent symmetry in nuclear spectroscopy. 
The occurrence of large deformations in the ground state and at low
excitations gives rise to rotational bands and large electric quadrupole transition intensities between states within the bands.  At higher excitations, much less is known experimentally.  Characterization of  this
part of the spectrum is needed for accurate calculation of the nuclear level density, which is very
sensitive to deformation and other structure
effects; observed level densities in rare-earth nuclei at the
neutron evaporation threshold vary by more than an order of
magnitude~\cite{RIPL}. In addition, nuclear fission is a phenomenon of shape
dynamics, and calculation of fission rates for excited nuclei requires
their level densities as a function of deformation~\cite{pe09}.

 Here we investigate nuclear deformation at finite temperature using the auxilliary-field Monte Carlo (AFMC) method,  which is well suited to the study of the evolution of nuclear properties with excitation energy while preserving rotational invariance.   In particular, we calculate the distribution of the quadrupole operator in the lab frame and demonstrate that it exhibits model-independent signatures of deformation.  We use moments of this distribution to calculate rotationally invariant  observables, which allow us to extract effective values of the intrinsic deformation and its fluctuations.  Deformations have been studied previously by the AFMC method, but with an ad hoc prescription to extract the intrinsic-frame properties~\cite{al96}.
 The methods presented here should be applicable to other finite-size systems in which correlations beyond the mean field are important. 

\ssec{Methodology} Formally, we can examine the statistical characteristics
of nuclei at finite excitations by calculating the thermal
expectation values of observables $\hat O$ associated with the property of
interest, $\langle {\hat O} \rangle = {\Tr\, ({\hat O} e^{-\beta \hat
H}) /\Tr\,  e^{-\beta \hat H}}$. 
Here $\beta^{-1}$ is the temperature and $\hat H$ is the Hamiltonian, which we assume to be rotationally invariant. 
We denote operators in the many-particle space
with a circumflex, to be distinguished from operators in the single-particle
space, which are ordinary matrices, denoted by
bold-face symbols.  Also, we denote the trace over the full many-particle
Fock space as $\Tr$ and the trace of matrices in the single-particle space
by $\tr$. The probability distribution of an operator $\hat O$,
$P_\beta(o)  ={ {\Tr (\delta[{\hat O} - o) e^{-\beta \hat H}]}/\Tr e^{-\beta \hat H}}$
 can be calculated using
the Fourier representation of the $\delta$ function:
\be
\label{prob}
P_\beta(o)={1\over \Tr\,  e^{-\beta \hat H}} \int_{-\infty}^\infty
{d \varphi \over 2 \pi} e^{-i \varphi o }\, \Tr\, \left(e^{i \varphi \hat O} e^{-\beta
\hat H} \right).
\ee 
Eq.~(\ref{prob}) is well-known for one-body observables $\hat O$ that
commute with the Hamiltonian, e.g., the number operator and the
$z$-component of the angular momentum~\cite{al07}.

Nuclear shape is different in that the relevant  operators, e.g., the
quadrupole operators, do not commute with the Hamiltonian.  Nevertheless, it
is possible with Eq.~(\ref{prob}) to define the distribution of
quantum-mechanical observables that carry information about deformation as
well as energy.  The distribution (\ref{prob}) can be
expressed in terms of the many-particle eigenstates of $\hat O$ and $\hat H$
as
\be\label{prob1}
P_\beta(o) = \sum_n \delta(o - o_n) \sum_m \langle o,n|e,m\rangle^2 e^{-\beta
e_m}.
\ee
Here $|o,n\rangle$ are eigenstates of $\hat O$ satisfying ${\hat
 O}|o,n\rangle = o_n |o,n\rangle$ and similarly for
$|e, m\rangle$.
Eq.~(\ref{prob1}) is valid whether or not the operators $\hat O$ and $\hat
H$ commute.  When they do commute, they share a common basis of eigenstates
such that $ \langle o,n|e,m\rangle=\delta_{m,n}$ and the distribution
(\ref{prob1}) reduces to its more familiar form $P_\beta(o) = \sum_n
\delta(o - o_n) e^{-\beta e_n}$. Note that in a finite model space the
eigenvalues $o_n$ form a discrete set and $P_\beta(o)$ is a finite sum of
$\delta$ functions.

In this work we consider the observable $\hat{O}$ to be the spectroscopic
mass quadrupole operator 
$\hat Q_{20} = \sum_i \left(2  z_i^2 -  x_i^2 - y_i^2 \right)$
where the sum is taken over all nucleons. The probability distribution
$P_\beta(q)$ of $\hat Q_{20}$  is defined
as in Eq.~(\ref{prob}) with $\hat O=\hat Q_{20}$ and $o=q$.

As we will show, this distribution can be accurately computed by the AFMC
method.  However, the intrinsic-frame properties are
not directly accessed by the operator $\hat Q_{20}$, which is a
laboratory-frame observable.  We shall demonstrate in this work that
nevertheless the distribution $P_\beta(q)$ is sensitive to deformation
effects and that the main properties of the deformation in the intrinsic
frame can be recovered from moments of this distribution.

Intrinsic frame quantities may be defined in terms of the expectation values
of rotationally invariant combinations of the quadrupole tensor operator
$\hat Q_{2\mu}$ ($\mu =-2, \ldots, 2$)~\cite{ku72,cl86}. The lowest-order invariant is
quadratic, $\hat Q\cdot \hat Q = \sum_\mu (-)^\mu \hat
Q_{2\mu} \hat Q_{2 -\mu}$.  There is one third-order invariant 
defined by coupling three quadrupole operators to angular momentum zero,
$(\hat Q \times \hat Q ) \cdot \hat Q = \sqrt{5} \sum_{\mu_1,\mu_2,\mu_3} \left( \begin{array}{ccc}
 2 & 2 & 2 \\ \mu_1 &  \mu_2 & \mu_3  \end{array}  \right) \hat Q_{2\mu_1}\hat Q_{2\mu_2}
\hat Q_{2\mu_3}$.
The fourth- and fifth-order invariants are also unique~\cite{Q-commutation} and we define
them as $ (\hat Q\cdot \hat Q)^2$ and $(\hat Q\cdot \hat Q)( (\hat Q \times
\hat Q ) \cdot \hat Q)$, respectively.  When the invariant  is unique at a given order, its expectation value can be
computed directly from the lab-frame moments of $\hat Q_{20}$, defined by 
$\langle \hat Q_{20}^n \rangle_\beta =  \int  q^n P_\beta(q)d q$.
The conversion factors are given in Table I.

\begin{table}[htb] 
\begin{center} 
\setlength{\tabcolsep}{.5em}
\begin{tabular}{|c|cccc|}
\hline 
n    &   2 &   3 &  4 & 5\\
\hline
invariant  &   5 & $-5(7/2)^{1/2}$ & 35/3 & $-(11/2)(7/2)^{1/2}$  \\
rotor & 1/5 & 2/35 & 3/35  &  4/77 \\ 
\hline 
\end{tabular} 
\caption{
First line: the ratio of the expectation value of the invariant of
order $n$ (see text) to the $n$-th moment of $\hat
Q_{20}$. Second line: the $n$-th moment of $\hat
Q_{20}$ for the rigid rotor in units of $q_0^n$ ($q_0$ is the rotor's intrinsic
quadrupole moment).
}
\end{center} 
\end{table}

\ssec{AFMC} We shall use the AFMC to evaluate the distribution in
Eq.~(\ref{prob}) for $\hat O=\hat Q_{20}$. AFMC is arguably the most
powerful computational tool for finding the ground states and thermal
properties in large-dimension many-particle spaces.  It is based on the
Hubbard-Stratonovich representation~\cite{HS} of the imaginary-time
propagator, $e^{-\beta \hat H} = \int D[\sigma] G_\sigma \hat U_\sigma$,
where $D[\sigma]$ is the integration measure, $G(\sigma)$ is a Gaussian
weight, and $\hat U_\sigma$ is a one-body propagator of non-interacting
nucleons moving in auxiliary fields $\sigma$. Practical
implementations require that the Hamiltonian be restricted to one- and
two-body terms, and that the two-body terms have the so-called good sign~\cite{sign}.  The method has been applied to nuclei in the framework of the configuration-interaction shell
model~\cite{la93,al94,SMMC}, where it is called the shell-model Monte
Carlo (SMMC).  It has been particularly successful in calculating
statistical properties of nuclei such as level densities~\cite{LD}. The
 distribution of $\hat Q_{20}$ is obtained from the Monte Carlo sampling
of fields $\sigma$ as a ratio of averages
\be\label{prob-HS}
P_\beta(q) = \left\langle \frac
{\Tr\left[\delta(\hat Q_{20} - q) \hat U_\sigma \right]}
{\Tr {\hat U}_\sigma} \Phi_\sigma \right\rangle_W 
 \langle \Phi_\sigma\rangle_W^{-1}  \;,
\ee
Here $\langle X \rangle_W =  \int D[\sigma] W_\sigma
X_\sigma/ \int D[\sigma] W_\sigma$, where $W_\sigma = G_\sigma |\Tr\, \hat
U_\sigma|$ is used for the Monte Carlo sampling  and $\Phi_\sigma=\Tr\, \hat U_\sigma/|\Tr \,\hat U_\sigma|$ is the
Monte Carlo sign function.

For a given $\hat U_\sigma$, we carry out the $\hat Q_{20}$ projection using
a discretized version of the Fourier decomposition in Eq.~(\ref{prob}). We
take an interval $[-q_{\rm max},q_{\rm max}]$ and divide it into $2M+1$ 
equal intervals of length $\Delta q=2q_{\rm
max}/(2M+1)$. We define $q_m=m \Delta q$, where
$m=-M,\ldots,M$, and approximate the quadrupole-projected trace in
(\ref{prob-HS}) by
\be\label{fourier-q}
\Tr\left(\delta(\hat Q_{20} - q_m) U_\sigma \right) \!\!  \approx \!\! {1\over 2 q_{\rm max}}\! \! \sum_{k=-M}^M
  \!\!\!  e^{-i \varphi_k q_m} \Tr(e^{i \varphi_k \hat Q_{20}} \hat U_\sigma) \;,
\ee
where $\varphi_k = \pi k/q_{\rm max}$ ($k=-M,\ldots, M$). Since $\hat
Q_{20}$ is a one-body operator and $\hat U_\sigma$ is a one-body propagator,
the Fock space many-particle traces on the r.h.s. of Eq.~(\ref{fourier-q})
reduce to determinants in the single-particle space 
$ \Tr\left(e^{i\varphi_k \hat Q_{20}} \hat U_\sigma \right) = 
  \det  \left( 1+   e^{i \varphi_k {\bf Q}_{20}} {\bf U}_\sigma\right)$.
Here ${\bf Q}_{20}$ and ${\bf U}_\sigma$ are  the matrices
representing, respectively, $\hat Q_{20}$ and $\hat U_\sigma$, in the
single-particle space.  In practice, projections are
carried on the neutron and proton number operators as well to fix the $Z$
and $N$ of the ensemble~\cite{SMMC}. 

 We found the thermalization of $\hat Q_{20}^n$
to be slow with the pure Metropolis sampling.  This can be overcome by
augmenting the Metropolis-generated configurations 
by rotating them through a properly chosen set of $N_{\Omega}$ rotation
angles  $\Omega$.  In practice, it is easier to
rotate the observables, i.e., we replace
$\langle e^{i \varphi \hat Q_{20}} \rangle_\sigma$ by ${1\over
N_\Omega} \sum_j \langle e^{i \varphi \hat Q_{20}(\Omega_j)} \rangle_\sigma$.
Here  $\hat
Q_{20}(\Omega)= \hat R \hat Q_{20} \hat R^{-1}$ with $\hat R$ being the
rotation operator for angle $\Omega$.  Details will be given elsewhere.  

We next discuss a few simple examples that can be treated
analytically or nearly so.

\begin{figure}[t]
  \includegraphics[angle=0,width=0.8\columnwidth]{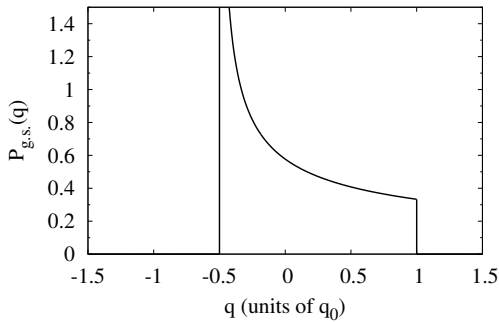}
    \caption{The ground-state distribution $P_{\rm g.s.}(q)$  vs $q/q_0$ for a prolate rotor with intrinsic quadrupole moment $q_0$.  }
    \label{rotor}
\end{figure}

\ssec{Rigid rotor} As a first simple example, we consider an axially
symmetric rigid rotor with an intrinsic quadrupole moment $q_0$ in its ground state.  The
distribution of its spectroscopic quadrupole operator in the laboratory
frame $Q_{20}= q_0(3 \cos^2 \theta -1)/ 2$ can be calculated in closed form.
For a prolate rotor ($q_0 > 0$)
\begin{eqnarray}\label{prob-rotor}
P_{\rm g.s.}(q) =  \left\{ \begin{array}{cc}  
\left(\sqrt{3} q_0 \sqrt{1 +2 {q\over q_0}}\right)  & \mbox{for $-{q_0 
\over 2} \leq q \leq q_0$}  \\
 0  & \mbox{otherwise}  \end{array}  \right. \;.
 \end{eqnarray}
This distribution is shown in Fig.~\ref{rotor}. The oblate rotor ($q_0 < 0$)
distribution is obtained from (\ref{prob-rotor}) by replacing $q$ with $-q$
and $q_0$ with $|q_0|$.
The moments of the distribution (\ref{prob-rotor}) can be calculated
from a simple recursion relation; their values for $2\leq n \leq 5$ are
given in Table I.  

\ssec {$^{20}$Ne} As a simple illustration in nuclear spectroscopy, we
consider the light deformed nucleus $^{20}$Ne.  The
orbital part of the single-particle wave functions are taken to be the
states of the $N=2$ harmonic oscillator shell, i.e., the $sd$-shell.   
The single-particle eigenvalues of ${\bf Q}_{20}$ are -2, 1, and 4 (in units of $b^2$~\cite{b}) with degeneracies 
of  6, 4 and 2, respectively.  The many-particle eigenvalues of $\hat
Q_{20}$ for $^{20}$Ne in the valence $sd$-shell
thus range from $-8$ to $16$  with a uniform spacing of
$3$.  The distribution $P_\beta(q)$ at $\beta=0$ is just  the
distribution of these eigenvalues. 

\begin{figure}[h!]
  \includegraphics[width=0.7\columnwidth]{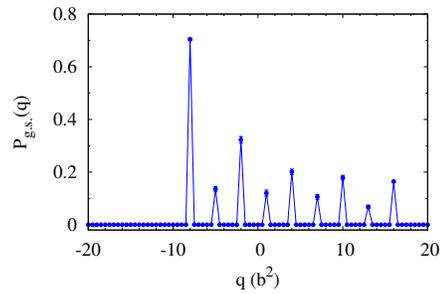}
    \caption{The AFMC ground-state quadrupole distribution $P_{\rm g.s.}(q)$  for $^{20}$Ne. The sharp $\delta$-like peaks demonstrate the discrete nature of the spectrum of $\hat Q_{20}$ and their envelope resembles the prolate rigid-rotor distribution in Fig.~\ref{rotor}. 
    \label{Ne20}}
\end{figure}

We have used this nucleus as a simple test of the AFMC. Here we
take the single-particle energies according to the USD
interaction~\cite{brown} and consider an attractive quadrupole-quadrupole
interaction $-\chi \tilde Q \cdot \tilde Q$, with $\tilde Q_{2\mu} = \sum_i
r_i^2 Y_{2 \mu}(\hat r_i)$  and 
$\chi = {8 \pi \over 5} {38.5\over A^{5/3}}$ MeV$/b^4$~\cite{la89}.
In Fig.~\ref{Ne20} we show the quadrupole distribution of the
$^{20}$Ne ground state. The discrete nature of the many-particle
eigenvalues of $\hat Q_{20}$ is evident; the distribution is a set $\delta$
functions at integers $-8,-5,\ldots,13,16$. The envelope of the strengths has
the skewed shape that looks qualitatively similar to the prolate rigid-rotor
distribution.

\ssec{SCMF}
It is instructive to compare
our results with those of the thermal SCMF,
e.g., the finite-temperature Hartree-Fock-Bogoliubov (HFB) approximation. 
The HFB solution is characterized by temperature-dependent one-body density matrix $\boldsymbol
\rho_\beta$ and pairing tensor $\boldsymbol \kappa_\beta$.  In general, two types of phase transitions can
occur vs temperature, a pairing transition and a deformed-to-spherical shape
transition~\cite{go86,ag00,ma03}. A
shape phase transition is also the generic result of a Landau 
theory in which the order parameter is the quadrupole deformation
tensor~\cite{al86}. The vast majority of deformed HFB  ground states are axially
symmetric~\cite{de10}, i.e.,  $\langle \hat
Q_{2\mu}\rangle = 0$ for $\mu \neq 0$. The second-order invariant
$\langle \hat Q\cdot \hat Q\rangle$ may be calculated in HFB by 
using Wick's theorem
\begin{eqnarray}\label{Q^2_HFB}
\langle \hat Q\cdot \hat Q\rangle = Q_0^2 + \sum_\mu (-)^\mu \tr\left[{\bf Q}_{2\mu}\, ({\bf 1}-\boldsymbol \rho_\beta)\, {\bf Q}_{2 -\mu}\, \boldsymbol \rho_\beta\right] \nonumber \\ + \sum_\mu (-)^\mu \tr\left[{\bf Q}_{2\mu}\, \boldsymbol \kappa_\beta \,{\bf Q}^T_{2 -\mu}\, \boldsymbol \kappa_\beta^\ast\right] \;,
\end{eqnarray}
where $ Q_0 \equiv \tr ({\bf Q}_{20} \boldsymbol \rho_\beta)$  is the intrinsic axial quadrupole moment.  The
remaining terms on the r.h.s.~of (\ref{Q^2_HFB}) represent the contributions due to quantal
and thermal fluctuations. 
We shall compare our AFMC results for rare-earth nuclei with the HFB theory in the next section.

\ssec{Rare-earth nuclei} Here we present results for rare-earth nuclei.  
The single-particle orbitals are taken from a Woods-Saxon potential plus
spin-orbit interaction; they span the $50-82$
shell plus $1f_{7/2}$ orbital for protons and the $82-126$ shell plus
$0h_{11/2}, 1g_{9/2}$ orbitals for neutrons.  We use the same interaction as
in Refs.~\cite{al08,oz13}.  The quadrupole moments are scaled
by a factor of 2 to account for the model space truncation.
\begin{figure}[h!]
  \includegraphics[width=0.7\columnwidth]{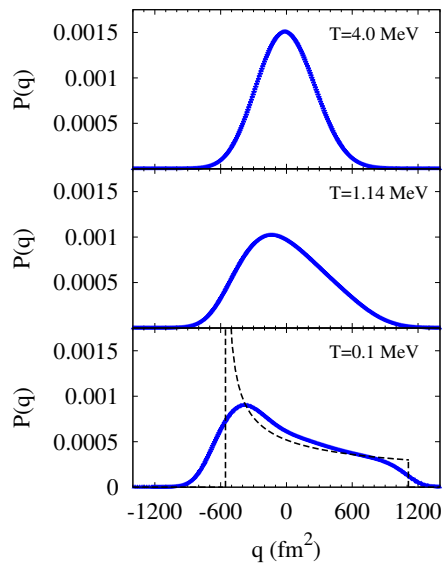}
    \caption{Probability distributions $P_\beta(q)$ for $^{154}$Sm at
 $T=0.1$ MeV,  $T=1.14$ MeV (shape transition temperature) and $T=4$ MeV.  
 The low-temperature distribution is compared with the rigid-rotor distribution (dashed line) and reflects the strongly deformed character of this nucleus. 
}   \label{P-T}
\end{figure} 

We first examine $^{154}$Sm, a strongly deformed nucleus with an
intrinsic quadrupole moment of $Q_0 \sim 1600$ fm$^2$, as determined
experimentally from in-band electric quadrupole transitions~\cite{ra01}.  AFMC $P_\beta(q)$ distributions are shown in Fig.~\ref{P-T} at three temperatures.  The distributions appear continuous because the many-particle eigenvalues of
$\hat Q_{20}$ are closely spaced.  
At the lowest temperature of $T=0.1$
MeV (bottom panel), $e^{-\beta \hat H}$ effectively projects out
the ground-state band. We observe the characteristic skewed distribution of
the prolate rotor.  The dashed line is the rotor distribution
(\ref{prob-rotor}) with $q_0$ taken at the HFB value of $Q_0$.  The middle panel is the distribution at the HFB shape
transition temperature, $T=1.14$ MeV. The
distribution is less skewed, but nevertheless it retains some trace of a prolate character.  The HFB excitation
energy at this temperature is about 25 MeV, much higher than energies of
interest for spectroscopy and for the neutron-capture reaction.  The top panel
shows the distribution at $T=4$ MeV.  At this high excitation the
distribution is featureless and close to a Gaussian.

We have also calculated $P_\beta(q)$
for $^{148}$Sm, which is spherical in its HFB ground state.  
They are more symmetric and 
change less with temperature, consistent with the absence of a coherent
quadrupole moment.
 
\begin{figure}[t]
  \includegraphics[width=0.8\columnwidth]{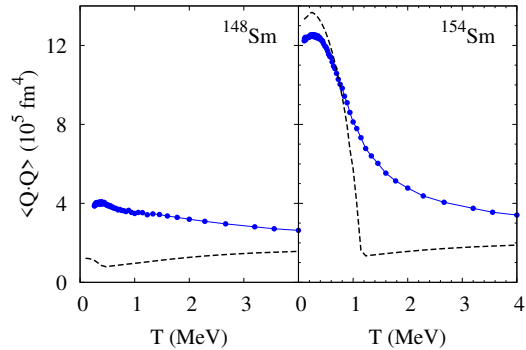}
    \caption{$\langle Q\cdot Q\rangle$ vs temperature $T$ for the spherical $^{148}$Sm (left) and the deformed $^{154}$Sm (right). The AFMC results (solid circles) are compared with the HFB results (dashed lines). }
    \label{Q^2_Sm}
\end{figure}

\ssec{Invariants}  Fig.~\ref{Q^2_Sm} shows the second-order invariant
$\langle \hat Q\cdot \hat Q\rangle$  vs temperature $T$  for $^{148}$Sm and $^{154}$Sm.  The AFMC
results (circles) are compared with the HFB
results (dashed lines) of Eq.~(\ref{Q^2_HFB}). In HFB, $\langle \hat Q\cdot \hat Q\rangle$ 
 for $^{148}$Sm can be entirely attributed to the
fluctuation terms in (\ref{Q^2_HFB}).  There is a small kink at $T=0.4$
MeV associated with the pairing transition, but by and large the curve
is flat.  The same is true of the AFMC curve.  In contrast, 
$\langle \hat Q\cdot \hat Q\rangle$ in $^{154}$Sm is very different at low temperatures. 
In HFB, the intrinsic quadrupole moment $Q_0$ is large, and it persists up
to a temperature of the order of $1$ MeV, close to the spherical-to-deformed
phase-transition temperature.  The AFMC results are in semiquantitative agreement at the lowest temperatures
showing that the coherent intrinsic quadrupole moment is not an artifact of
the HFB.  The sharp kink characterizing the HFB shape
transition~\cite{go86,ma03} is washed out, as is expected in a
finite-size system. Nevertheless a signature of this phase transition
remains in the rapid decrease of $\langle Q\cdot Q\rangle$ with temperature.
In AFMC deformation effects survive
well above the transition temperature, in that $\langle Q\cdot Q\rangle$
continues to be enhanced beyond its uncorrelated mean-field
value.

The second- and third-order invariants can be used to define effective values of the intrinsic
shape parameters $\beta,\gamma$~\cite{beta} of the collective Bohr
model~\cite[Sec. 6B-1a]{BM}. The model assumes an intrinsic frame in which
the quadrupole deformation parameters $\alpha_{2\mu} = \sqrt{5 \pi}\langle \hat
Q_{2\mu}\rangle
/3 r_0^2 A^{5/3}$ are expressed as $\alpha_{20}=\beta\cos \gamma$,
$\alpha_{22}=\alpha_{2-2}=\frac{1}{\sqrt{2}}\beta\sin\gamma$, and
$\alpha_{2\pm1}=0$.  Effective $\beta$ and $\gamma$ can then be
determined from the corresponding invariants
\be
\beta = \frac{\sqrt{5 \pi}}{ 3 r_0^2 A^{5/3} } \langle \hat Q \cdot \hat Q \rangle^{1/2}  \;;\;\;
\cos 3\gamma = -\sqrt{7 \over 2} {\langle (\hat Q \times \hat Q ) \cdot \hat Q \rangle \over  \langle \hat Q \cdot \hat Q \rangle^{3/2} } \;.
\ee
In addition, we can extract a 
measure $\Delta \beta$ of the fluctuations in $\beta$ 
using the second- and fourth-order invariants
\be
\left({\Delta \beta / \beta}\right)^2 =  {\left[\langle (\hat Q \cdot \hat Q)^2 \rangle - \langle \hat Q \cdot \hat Q \rangle^2\right]^{1/2} / \langle \hat Q \cdot \hat Q \rangle} \;.
\ee
The invariants themselves are calculated from the moments of $P_\beta(q)$
using the relations in Table I. 
As expected, the deformed $^{154}$Sm has a
larger deformation $\beta$ than $^{148}$Sm ($0.232$ vs $0.137$), but a
smaller deformation angle $\gamma$ ($13.4^{\circ}$ vs $21.6^{\circ}$)
that is closer to an axial shape.  The deformed nucleus is more rigid in
that it has a smaller $\Delta \beta/\beta$, $0.51$ for $^{154}$Sm vs
$0.72$ for $^{148}$Sm.

\ssec{Summary}  We have demonstrated that the distribution of
the axial quadrupole operator can be computed in the AFMC method, and that it
conveys important information about deformation and the intrinsic shapes of
nuclei at finite temperature.  In particular, the expectation values of
$\beta^2$, $\beta^3 \cos 3\gamma$ and the fluctuation in $\beta^2$ can be
extracted as a function of temperature.  With these moments, it should be possible to construct models of the joint level density
distribution $\rho(\beta,E_x) = \rho(E_x)
P_{E_x}(\beta)$, where $\rho(E_x)$ is the total level density and $
P_{E_x}(\beta)$ is the intrinsic shape distribution at excitation
energy $E_x$.  This joint distribution is an important component in
the theory of fission and will be discussed in a future publication.

\ssec{Acknowledgments} 
Y. Alhassid acknowledges the hospitality of the Institute of Nuclear Theory
in Seattle, where part of this work was completed.  We thank H. Nakada for the HFB code. This work was supported
in part by the U.S. DOE grant Nos. DE-FG02-91ER40608 and DE-FG02-00ER411132.
Computational
cycles were provided by the NERSC high performance computing facility 
and by the facilities of the Yale University Faculty of Arts and Sciences
High Performance Computing Center.


\begin{thebibliography}{99}
\bibitem{su3}  The $SU(3)$ representation provides an alternate way
to study deformation effects~\cite{va02}, but it has not been widely
applied to heavy nuclei.
\bibitem{Q-commutation} Strictly speaking the uniqueness of the fourth- and fifth-order invariants holds when the operators $\hat Q_{2\mu}$ commute among themselves.  The quadrupole operators commute in coordinate space but generally not in a truncated shell model space.  However, the effects due to their non-commutation is small.
\bibitem{sign} Small bad-sign interaction terms can be treated using the extrapolation method of Ref.~\cite{al94}.
\bibitem{b} The oscillator length parameter $b$ is defined to 
give a ground-state wave function of the form $ \sim \exp(-r^2/2b^2)$.
\bibitem{beta} We use the symbol $\beta$ for both the inverse temperature and
the dimensionless deformation parameter; the meaning should be clear from
the context.
\end{thebibliography}
\end{document}